%% file: ADC_compute_work.tex
\documentclass[conference]{IEEEtran}
\IEEEoverridecommandlockouts
\usepackage{cite}
\usepackage{amsmath,amssymb,amsfonts}
\usepackage{algorithmic}
\usepackage{graphicx}
\usepackage{textcomp}
\usepackage{xcolor}
\usepackage[acronym]{glossaries}
\usepackage{subfig}
\usepackage{colortbl}
\usepackage{makecell}
\usepackage{multirow}
\usepackage[hidelinks]{hyperref}
\usepackage{vcell}
\usepackage[capitalise]{cleveref}
\usepackage{url}
\usepackage{units}
\usepackage{threeparttable}
\usepackage{booktabs}
\usepackage{colortbl}
\usepackage{tabularx}
\usepackage{makecell}
\usepackage{tikz}

\def\BibTeX{{\rm B\kern-.05em{\sc i\kern-.025em b}\kern-.08em
    T\kern-.1667em\lower.7ex\hbox{E}\kern-.125emX}}

\input{ieee_cols.tex}

\newcommand*\circnum[1]{\tikz[baseline=(char.base)]{%
            \node[white,shape=circle,fill=ieee-dark-black-100,draw,inner sep=1pt] (char) {\color{ieee-bright-white-100}\sffamily #1};}}

\newcommand*\circnumsmall[1]{\tikz[baseline=(char.base)]{%
            \node[white,shape=circle,fill=ieee-dark-black-100,draw,inner sep=1pt] (char) {\small\color{ieee-bright-white-100}\sffamily #1};}}

\renewcommand{\baselinestretch}{0.96}
\usepackage{lettrine}
\usepackage{orcidlink}

\begin{document}

\newacronym{aimc}{AIMC}{Analog In-Memory Computing}
\newacronym{ai}{AI}{Artificial Intelligence}
\newacronym{adc}{ADC}{Analog-to-Digital Converter}
\newacronym{bn}{BN}{Batch Normalization}
\newacronym{cnn}{CNN}{Convolutional Neural Network}
\newacronym{cv}{CV}{Coefficient of Variation}
\newacronym{dac}{DAC}{Digital-to-Analog Converter}
\newacronym{dl}{DL}{Deep Learning}
\newacronym{dnn}{DNN}{Deep Neural Network}
\newacronym{fp16}{FP16}{Floating Point 16}
\newacronym{fp32}{FP32}{Floating Point 32}
\newacronym{gpu}{GPU}{Graphic Processing Unit}
\newacronym{imc}{IMC}{In-Memory Computing}
\newacronym{ldpu}{LDPU}{Local Digital Processing Unit}
\newacronym{lsb}{LSB}{Least Significant Bit}
\newacronym{ml}{ML}{Machine Learning}
\newacronym{msb}{MSB}{Most Significant Bit}
\newacronym{mvm}{MVM}{Matrix-Vector-Multiplication}
\newacronym{nmpu}{NMPU}{Near-Memory digital Processing Unit}
\newacronym{pcm}{PCM}{Phase Change Memory}
\newacronym{ppa}{PPA}{power, performance and area}
\newacronym{relu}{ReLU}{Rectified Linear Unit}
\newacronym{rtl}{RTL}{Register Transfer Level}

\vspace{-0.3cm}
\title{A Precision-Optimized Fixed-Point Near-Memory Digital Processing Unit for Analog In-Memory Computing \vspace{-0.25cm}}
\author{\IEEEauthorblockN{Elena Ferro\orcidlink{0000-0002-8618-8643}\thanks{\hspace{-1em}\rule{3cm}{0.5pt} \newline \textcopyright  \hspace{1pt} 2024 IEEE. Personal use of this material is permitted. Permission from IEEE must be obtained for all other uses, in any current or future media, including reprinting/republishing this material for advertising or promotional purposes, creating new collective works, for resale or redistribution to servers or lists, or reuse of any copyrighted component of this work in other works.}\IEEEauthorrefmark{1}\IEEEauthorrefmark{3}\IEEEauthorrefmark{2}, Athanasios Vasilopoulos\orcidlink{0009-0001-9081-6139}\IEEEauthorrefmark{2}, Corey Lammie\orcidlink{0000-0001-5564-1356}\IEEEauthorrefmark{2}, Manuel Le Gallo\orcidlink{0000-0003-1600-6151}\IEEEauthorrefmark{2},\\
Luca Benini\orcidlink{0000-0001-8068-3806}\IEEEauthorrefmark{3}, Irem Boybat\orcidlink{0000-0002-4255-8622}\IEEEauthorrefmark{1}\IEEEauthorrefmark{2}, Abu Sebastian\orcidlink{0000-0001-5603-5243}\IEEEauthorrefmark{2}}
\IEEEauthorblockA{\IEEEauthorrefmark{2} IBM Research Europe, 8803 R\"{u}schlikon, Switzerland,\IEEEauthorrefmark{3} IIS-ETH Z\"{u}rich, 8092 Z\"{u}rich, Switzerland\\
\IEEEauthorrefmark{1}Email: \{elf, ibo\}@zurich.ibm.com}}
\IEEEaftertitletext{\vspace{-1.5\baselineskip}}
\maketitle
\begin{abstract}
\gls{aimc} is an emerging technology for fast and energy-efficient \gls{dl} inference. However, a certain amount of digital post-processing is required to deal with circuit mismatches and non-idealities associated with the memory devices. Efficient near-memory digital logic is critical to retain the high area/energy efficiency and low latency of \gls{aimc}.
Existing systems adopt \gls{fp16} arithmetic with limited parallelization capability and high latency. To overcome these limitations, we propose a \gls{nmpu}  based on fixed-point arithmetic. It achieves competitive accuracy and higher computing throughput than previous approaches while minimizing the area overhead. Moreover, the \gls{nmpu} supports standard \gls{dl} activation steps, such as ReLU and Batch Normalization.
We perform a physical implementation of the \gls{nmpu} design in a \unit[14]{nm} CMOS technology and provide detailed performance, power, and area assessments.
We validate the efficacy of the \gls{nmpu} by using data from an AIMC chip and demonstrate that a simulated \gls{aimc} system with the proposed \gls{nmpu} outperforms existing \gls{fp16}-based implementations, providing \unit[139]{$\times$} speed-up, \unit[7.8]{$\times$} smaller area, and a competitive power consumption. 
Additionally, our approach achieves an inference accuracy of \unit[86.65]{\%}/\unit[65.06]{\%}, with an accuracy drop of just \unit[0.12]{\%}/\unit[0.4]{\%} compared to the \gls{fp16} baseline when benchmarked with ResNet9/ResNet32 networks trained on the CIFAR10/CIFAR100 datasets, respectively.
 \end{abstract}

\begin{IEEEkeywords}
Near-memory processing, Fixed-point computing, Analog in-memory computing, Deep learning, AI
\end{IEEEkeywords}

\glsresetall

\section{Introduction}
The recent growth in the \gls{ai} domain has been driven by the shift from compute-centric systems to data-centric computing systems~\cite{Siegl_2016}.
However, conventional von Neumann architectures exploiting data-level parallelism are bound by the \emph{memory-wall}, as well as the microarchitectural challenges imposed by the increasing degree of concurrency required, which affect latency and energy efficiency~\cite{Keckler_2011, Jouppi_2017, Sze_2017}. 
To overcome these limitations, a promising alternative leverages non-von Neumann architectures, which perform computational tasks within the memory itself. 
This paradigm, known as \gls{imc}, tackles the von Neumann bottleneck, leading to significant improvements in both energy efficiency and latency. 
A promising realization of \gls{imc} is \gls{aimc}, with recent works showcasing its potential~\cite{Khaddam_2022,Gallo_2023,Wan_2022}. 
The \gls{mvm} operation, which dominates the computations ($>$\unit[95]{\%}) of \gls{dnn} inference, can be performed by mapping offline-trained network weights onto \gls{aimc} tiles. 
Consequently, \gls{aimc} is a very promising approach for accelerating these workloads~\cite{Sebastian_2020}. 
For example, a single 256$\times$256 \gls{pcm} tile has been shown to achieve a peak energy efficiency of $\sim$\unit[10]{TOPS/W}~\cite{Khaddam_2022,Gallo_2023} while innovative design approaches are aiming for $\sim$\unit[100]{TOPS/W}~\cite{Jain_2023}.

\begin{figure}[t]
\centerline{\includegraphics[width=0.8\columnwidth]{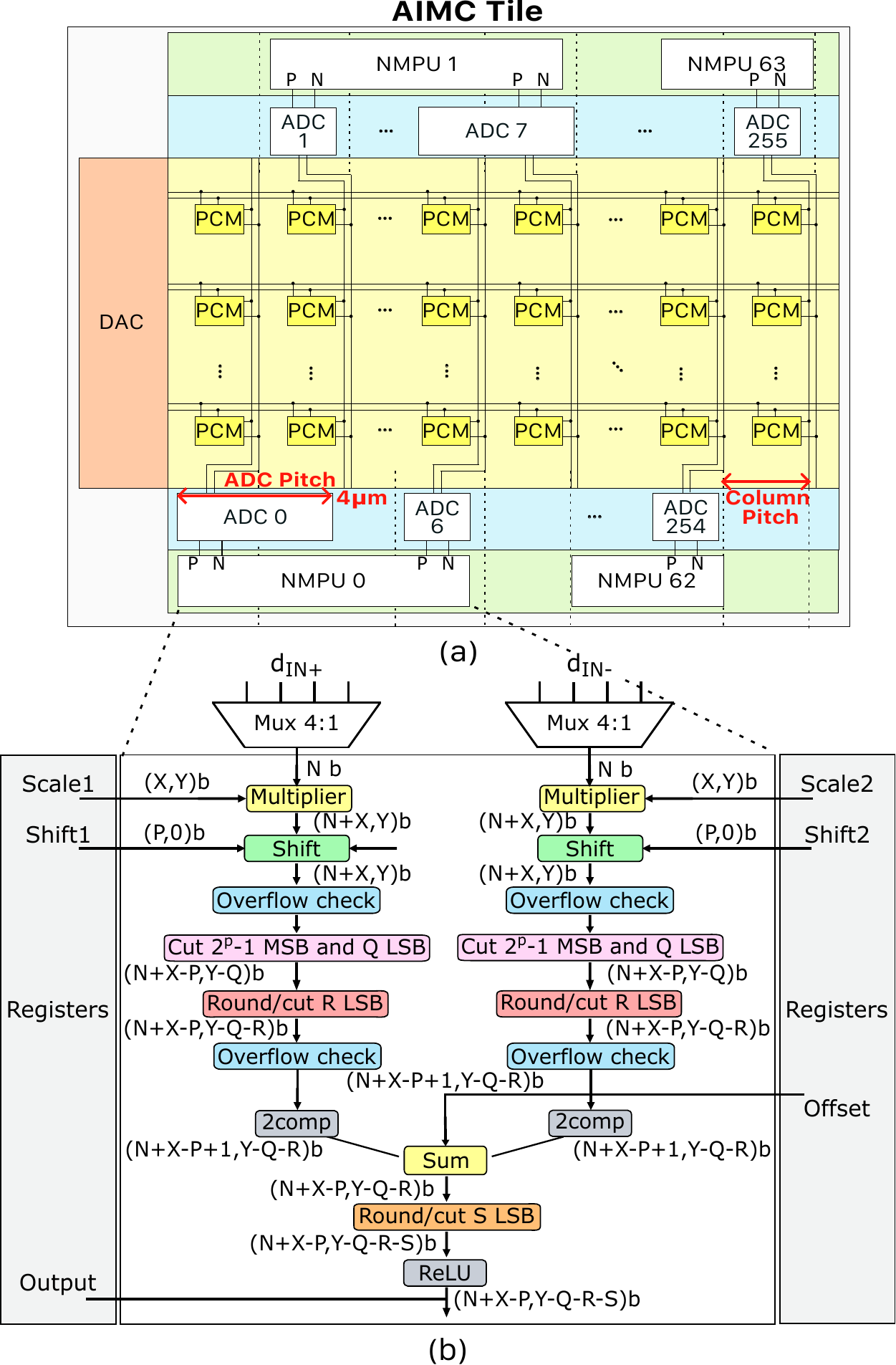}}
\caption{(a) Sample \gls{aimc} tile consisting of a 256$\times$256 PCM crossbar, DAC block, 256 ADCs and 64 NMPUs; (b) flow diagram of the proposed fixed-point NMPU.}
\label{fig:aimc_system}
\end{figure}
A typical \gls{aimc}-tile comprises a crossbar array of memristive devices, such as \gls{pcm}. The synaptic weights are programmed as the conductance value of these devices. The digital input to the tile is converted to voltage or pulse duration values using \glspl{dac} or pulse-width modulation units, respectively, and the output from the array is converted to digital values using \glspl{adc}. However, one key challenge is that there is substantial non-uniformity in the \glspl{adc}' conversion behavior (or transfer curves). To address this, \gls{aimc} tiles require additional data post-processing per column through an \emph{affine correction} procedure computed as:
\setlength{\belowdisplayskip}{2.5pt}
\setlength{\abovedisplayskip}{2.pt}
\begin{equation}
    d_{\text{out},\text{aff}} = d_{\text{in}} \times \text{scale}_{\text{aff}} + \text{offset}_{\text{aff}},    \label{eq:affine_correction}
\end{equation}
where $d_{\text{in}}$ is the output of the \gls{adc}, and $\text{scale}_{\text{aff}}$ and $\text{offset}_{\text{aff}}$ are the affine correction parameters.
These parameters can also be leveraged to perform any subsequent scaling and offset operation on $d_{\text{out},\text{aff}}$. For instance, \gls{pcm} devices exhibit significant temporal conductance drift and the conductance values also vary with temperature. These effects can be partially compensated by using correction factors~\cite{Nandakumar_2020,Boybat_2021}. Moreover, the affine correction parameters can also realize \gls{bn} when implementing \glspl{dnn} such as \glspl{cnn}. Hence, by providing additional support for the \gls{relu} activation function, the \gls{aimc} tile periphery provides compatibility with \gls{cnn} auxiliary operations. Therefore, we formulate the operations to be performed by near-memory processing block targeting \gls{cnn} implementations as follows:  
\begin{equation}
    d_{\text{out}} = \text{ReLU}(d_{\text{in}} \times \text{scale} + \text{offset}).    \label{eq:affine_correction2}
\end{equation}

Clearly, near-memory digital processing is an indispensable part of an AIMC tile. 
To achieve high computational efficiency and low latency in \gls{aimc} systems, it is desirable to integrate the custom near-memory digital logic circuitry for each crossbar column. 
However, maximizing the area efficiency of the tile (TOPS/mm$^2$), as well as complying with the small pitch of the \gls{aimc} columns and the \glspl{adc} (for instance, \unit[4]{$\mu$m} in~\cite{Khaddam_2022}), heavily constraints the available area budget and the physical implementation of this block (\cref{fig:aimc_system}a). 
Therefore, a flexible solution maintaining high energy efficiency and accuracy while being compact and fast is required. Designs using \gls{fp16} datatypes were explored in prior works~\cite{Khaddam_2022, Gallo_2023}. However, these approaches incur a relatively large area and power consumption. This leads to a large digital block shared across multiple columns (e.g., 256 columns in~\cite{Gallo_2023}), which limits parallelization and requires serial pipelined processing for all outputs. 
An alternative is the adoption of standard integer types, which reduce the area overhead and enable multiplexing across 8~\cite{Jia_2022} or 16~\cite{Jia_2020} crossbar columns.

In this paper, we address the limitations of existing solutions by proposing a flexible \gls{nmpu} that leverages fixed-point precision computation to perform accurate affine correction and auxiliary operations while minimizing its area footprint. Our design can be multiplexed across just 4 memristive crossbar array columns and, compared to~\cite{Jia_2022,Jia_2020}, adopts truncation and rounding schemes, significantly reducing the size of intermediate representations for a more compact design.
We experimentally validate the effectiveness of our approach using \gls{adc} data from a \gls{pcm}-based \gls{aimc} chip~\cite{Khaddam_2022}. 

\section{NMPU microarchitecture and design space exploration}
The \gls{nmpu} microarchitecture implementing the computations of \eqref{eq:affine_correction2} is shown in \cref{fig:aimc_system}b.
The design consists of two identical branches, which apply different scale parameters to the digitized values of positive and negative currents coming from one \gls{adc} as distinct paths~\cite{Khaddam_2022, Gallo_2023}. 
The block consumes two 10-bit unsigned integer inputs and produces an 8-bit signed output. 
In addition, the unit implements rounding and truncation steps to reduce intermediate precision and mitigate precision loss during the computations. 
The design is fully parameterizable in terms of the inputs, scale, shift and offset factors, intermediate truncation formats, and output format.
In the following, we explore optimal parameters for the proposed block. For this analysis, we set the input and output formats as 10-bit unsigned (integer) and 8-bit signed (2's complement), respectively. The choice of the input format is based on the adoption of a 10-bit \glspl{adc}' design, while 8-bit outputs for activations are widely adopted for \gls{dl} inference acceleration~\cite{TPU_2017, Sze_2017, Sze_2019, The_next_platform_2016, Reuther_2022}.

\begin{figure}[t]
\centerline{\includegraphics[width=.85\columnwidth]{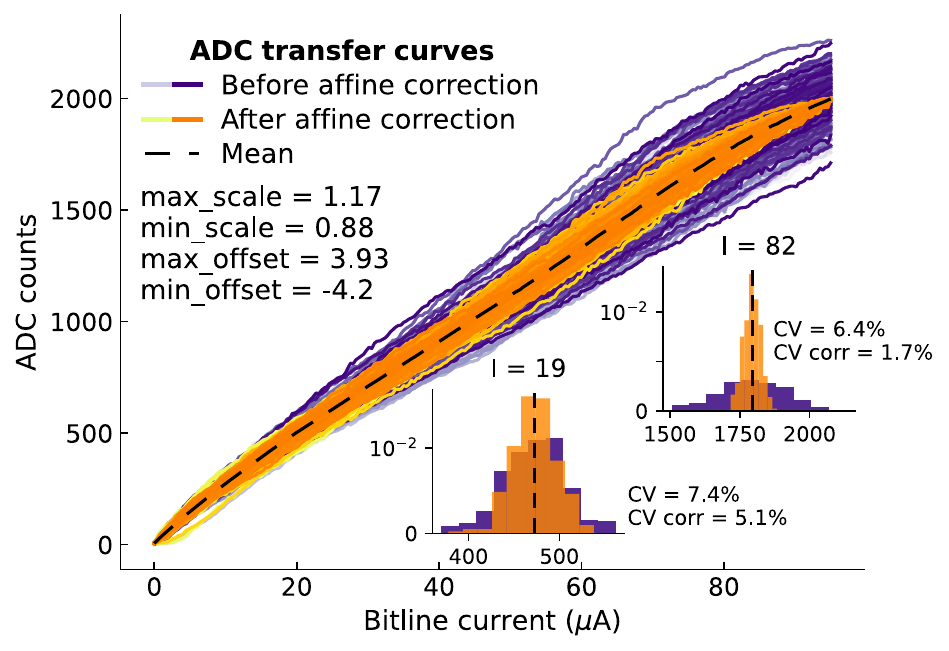}}
\caption{Set of 256 \gls{adc} transfer curves: after calibration in purple; and after the affine correction in orange.}
\label{fig:ADC_transfer_curve}
\vspace{-0.12cm}
\end{figure}

First, we determine the optimal data formats for scale and offset constants by analyzing the \gls{adc} transfer curves, which are experimentally acquired from circuit measurements of~\cite{Khaddam_2022} and shown in \cref{fig:ADC_transfer_curve}. %
They are obtained for a set of 256 distinct \glspl{adc} after being properly calibrated~\cite{Khaddam_2022}.
The individual trajectories highlight the variations across individual \glspl{adc}, and the relationship between the bit-line current and the converted \gls{adc} counts is observed to be nonlinear. 
The affine correction (see (\ref{eq:affine_correction})) minimizes the variation of the curves with respect to their mean, i.e., tuning the slope with a certain precision granularity. 
We estimate the maximum and minimum scaling factors for the affine correction as 0.88 and 1.17, respectively. 
Hence, 1-bit for the integer part is sufficient in the corresponding fixed-point number.
On the other hand, the fractional part depends on the desired precision for the tuning, which is targeted to be higher than the precision of the \gls{cv} in the set of 256 \glspl{adc} transfer curves. 
Since the \gls{cv} is observed to be $\simeq$\unit[7]{\%} of the mean, we retain a granularity of \unit[1]{\%}, which corresponds to $\simeq2^{-7}$, i.e., 7-bit.
Thus, we need 1-bit and 7-bit to represent the integer and fractional part of the scale parameter, respectively, i.e., (1,7) bit unsigned. 
For the offset parameter, we observe that (4,0) bit signed is sufficient to account for the measured offset.
However, we allocate more bits to accommodate for any further requirement associated with the \gls{bn} operation and propose using (7,1) bit signed.

To perform the scaling in high precision and to ensure that the scaled values fit the output range, we introduce a \textit{right shift} operation after the scale (see \cref{fig:aimc_system}b). 
Given that the \glspl{adc}' output precision is 10-bit unsigned, the scale is 1-bit precision for the integer part, and the output is 8-bit signed, the maximum value needed for the right shift is three, which can be represented with 2 bits.
In addition, since the shift operation ensures the range fits 8-bit, we reduce the integer bits by cutting the unused \glspl{msb}, i.e., ($2^2-1$) \glspl{msb}.
Furthermore, since a high degree of precision is unnecessary when determining the methodology for conducting the rounding operation, we truncate the \glspl{lsb} up to and including $2^{-6}$.
\begin{figure}[t]
\centerline{\includegraphics[width=.7\columnwidth]{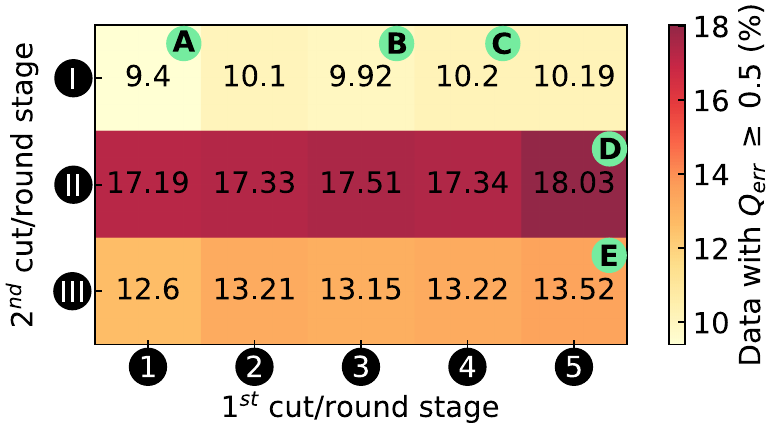}}
\caption{Quantization error for the 15 explored \glspl{nmpu} adopting different first and second cut/round stages.}
\label{fig:quantization_comparison}
\vspace{-0.09cm}
\end{figure}

To evaluate the influence of cut and round operations on the overall precision loss, we consider 15 distinct architectures leveraging a different combination of cut/round methods after the scaling (\textit{first cut/round stage}) and the sum (\textit{second cut/round stage}).
We denote as guard (\text{G}) and round (\text{R}) the bits that determine whether a truncation or rounding should occur, respectively, and as \text{x} the bits not affecting the decision.
For the first cut/round stage, we consider five methods: \circnum{1} cut 3 \glspl{lsb} up to $2^{-2}$ (excluded) and round only looking at the bit at position $2^{-3}$ (x.xx$\vert$Rxx); \circnum{2} cut 3 \glspl{lsb} up to $2^{-2}$ (excluded) and if the bit at position $2^{-2}$ is 1 then we cut; otherwise, we round looking at the bits at position $2^{-3}$ or $2^{-4}$ (if at least one is 1) (x.xG$\vert$RRx); \circnum{3} cut 3 \glspl{lsb} up to $2^{-2}$ (excluded) and if the bit at position $2^{-2}$ is 1 then we cut; otherwise, we round looking at the bit at position $2^{-3}$ (x.xG$\vert$Rxx); \circnum{4} cut 3 \glspl{lsb} up to $2^{-2}$ (excluded) and if the bit at position $2^{-2}$ is 1 then we cut; otherwise, we round looking at the bit at position $2^{-3}$ or $2^{-4}$ or $2^{-5}$ (x.xG$\vert$RRR); \circnum{5} cut 3 \glspl{lsb} up to $2^{-2}$ (excluded) (x.xx$\vert$xxx).

For each of these methods, we consider three ways of performing the second cut/round stage: \circnum{I} cutting the \gls{lsb} for both positive and negative values; \circnumsmall{II} rounding the \gls{lsb} for positive and cutting \gls{lsb} for negative values; and \circnumsmall{III} rounding \gls{lsb} for both positive and negative values.

\section{Evaluating precision of the proposed datapath}      
To evaluate the impact of the two cut/round stages on the precision loss, we evaluate the architectures with a cycle-accurate \gls{rtl} simulator, using randomly generated data from a univariate normal Gaussian distribution with mean 0 and variance 1 for the inputs. 
As an evaluation metric for the precision loss, we compute the quantization error against a software-based (FP32) baseline:
\begin{equation}
    Q_{\text{err}} = |\text{out\_hardware} - \text{out\_baseline}|
\end{equation}
\begin{figure}[t]
\center
\centerline{\includegraphics[width=.88\columnwidth]{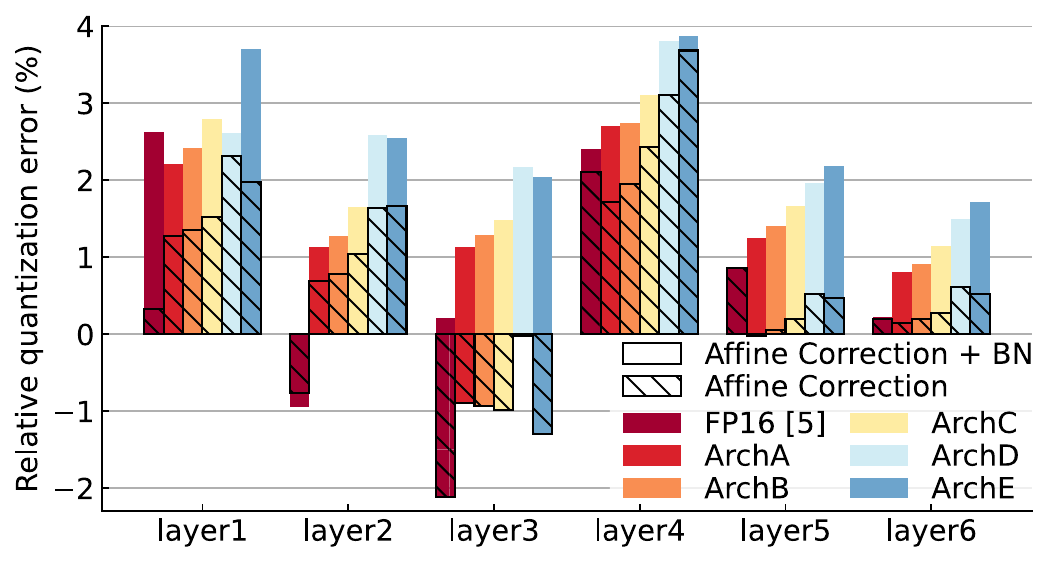}}
\caption{Relative quantization error on ResNet9 layers.}
\label{fig:accuracy_comparison_RealData}
\vspace{-0.13cm}
\end{figure}
Being the output integer, the best architectures provide the lowest percentage of data with $Q_{\text{err}} \geq 0.5$, as shown in \cref{fig:quantization_comparison}. 
We observe that $Q_{\text{err}}$ is mainly affected by the method adopted for the \emph{second cut/round stage}. 
In particular, the architectures implemented with \circnum{I} have less than \unit[11]{\%} of data (over a range of 10,000 data points) with $Q_{\text{err}} \ge 0.5$. 
We continue the analysis by selecting the two most performant architectures (\textit{A} and \textit{B}), as well as the least performant for each method in the \textit{second cut/round stage} (\textit{C}, \textit{D} and \textit{E}), as shown in \cref{fig:quantization_comparison}.

Next, we implement Python models for the selected architectures and evaluate them using \gls{adc} data from the chip presented in~\cite{Khaddam_2022}. Data is extracted during inference on a modified ResNet9 model~\cite{Buechel_2022}.
We compare the relative quantization error on a layer-by-layer basis (i.e., without any layer propagation) with the software baseline.
More specifically, the error per layer is evaluated as follows. 
First, we compute the \gls{mvm} and, where applicable, the \gls{bn}, all in FP32, to serve as a software baseline.
Second, we estimate the hardware operation error (``hw\_op\_err'') resulting from performing both \gls{mvm} and analog-to-digital conversion on chip when the affine correction and \gls{bn} are performed in FP32 (\gls{aimc}-baseline).
This hardware operation error is computed as the root-mean-squared (L2) error between the \gls{aimc}-baseline and the software baseline results.
Then, we estimate the L2-error between the software baseline and the proposed implementations (referred to as ``impl\_err''), where the \gls{mvm} is performed on-chip and the affine correction and \gls{bn} are performed with the selected \gls{nmpu} variants.
Finally, the relative quantization error is calculated as:
\begin{equation}
    Q_{\text{err\_rel}} = \frac{\text{impl\_err - hw\_op\_err}}{\text{hw\_op\_err}}.
\end{equation} 
\cref{fig:accuracy_comparison_RealData} shows that the \gls{nmpu} block incurs a relative quantization error $<$\unit[4]{\%} and $<$\unit[2]{\%} compared to the \gls{aimc}-baseline and to the \gls{fp16}-based post-processing implementation from ~\cite{Khaddam_2022} for all layers, respectively. 
When including \gls{bn}, $Q_{\text{err\_rel}}$ increases for both the \gls{nmpu} and the \gls{fp16} compared to the \gls{aimc}-baseline.
In both situations, \gls{nmpu}'s $Q_{\text{err\_rel}}$ varies with a trend similar to that of \cref{fig:quantization_comparison}. In the best case, i.e., ArchA, the \gls{nmpu} achieves a $Q_{\text{err\_rel}}<$\unit[1]{\%} compared to \gls{fp16}. 
Consequently, in circuit applications, where the absolute value on the output is crucial, exploring different cut/round methods is important to identifying the architecture with the lowest output error.
However, when targeting \gls{dl} applications, the final goal is the accuracy assessment rather than comparing absolute output values.
Therefore, in the following, we investigate the impact of the quantization error on the accuracy evaluation, building towards a more comprehensive application study.
\begin{figure}[t]
\centerline{\includegraphics[width=.95\linewidth]{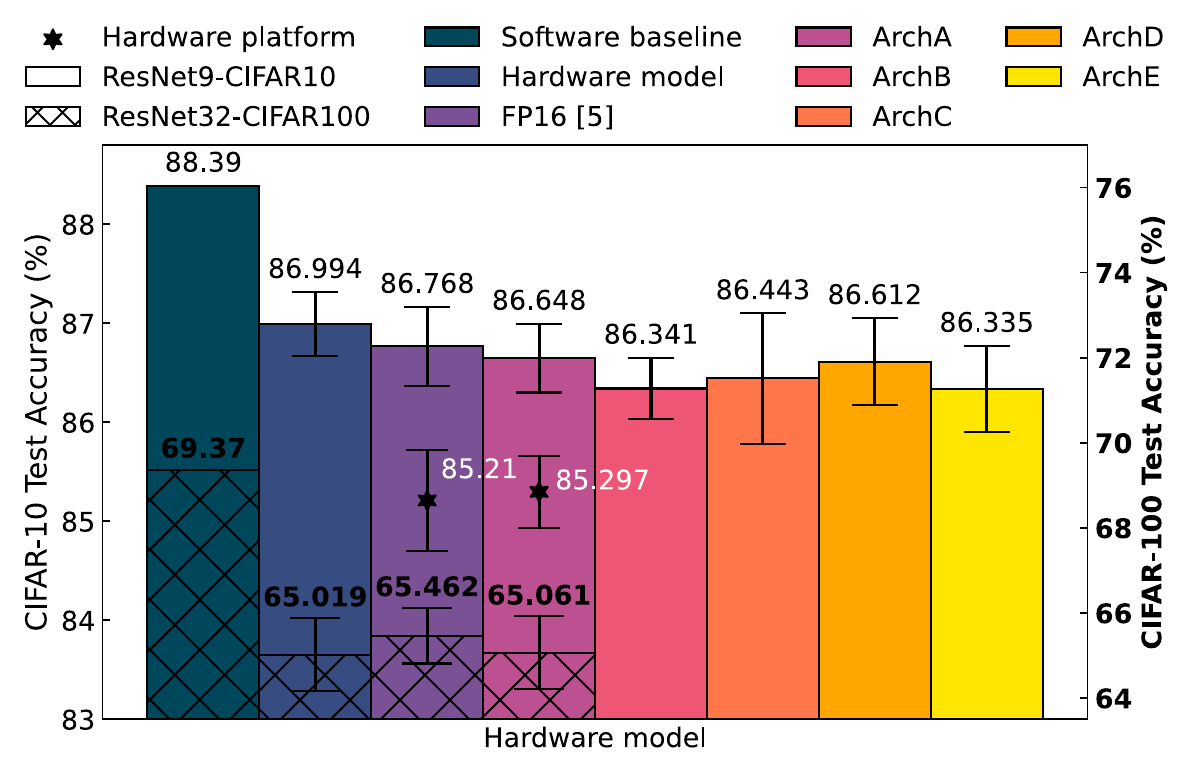}}
\caption{Inference accuracy for ResNet9 and ResNet32 for different architectures. The error bars represent the standard deviation obtained from 10 repetitions.}
\label{fig:accuracy_comparison_resnet9}
\vspace{-0.06cm}
\end{figure}

Besides the previous evaluation, we feed the output of the near-memory post-processing block to the \gls{aimc} tile containing the weights of the subsequent layer. 
For this study, we simulate the hardware using statistical models extracted from chip measurements~\cite{Gallo_2023} and explore both ResNet9 and a deeper ResNet32~\cite{He_2016} on the CIFAR10 and CIFAR100 datasets, respectively. 
\cref{fig:accuracy_comparison_resnet9} shows the results obtained via simulations. We observe an accuracy degradation from the software baseline (similar to that of \cref{fig:accuracy_comparison_RealData}) when the hardware model is used. 
The hardware model employs statistical models of the chip to simulate the \gls{mvm} and data converters, while the affine correction and \gls{bn} are performed in FP32. 
In the other implementations from \cref{fig:accuracy_comparison_resnet9}, the hardware model is revised to perform the affine correction and \gls{bn} using \gls{fp16}~\cite{Khaddam_2022} or the proposed \glspl{nmpu}.
For ResNet9 on the CIFAR10 dataset, the selected architectures show an inference accuracy drop $\le$~\unit[2.1]{\%} compared to the software baseline, $\le$~\unit[0.7]{\%} compared to the hardware model with an FP32 periphery, and $\le$~\unit[0.5]{\%} compared to the \gls{fp16}~\cite{Khaddam_2022} model.
Overall, independently of the observed $Q_\text{err}$, the fixed-point-based \glspl{nmpu} do not significantly affect the inference accuracy, whose degradation is instead dominated by the \gls{mvm} non-idealities. 

Since the accuracy of the explored architectures is similar, for ResNet32 on the CIFAR100 dataset and for the following explorations, we consider only the architecture with the highest inference accuracy mean, i.e. \unit[86.648]{\%}, referred to as \textit{ArchA}.
\cref{fig:accuracy_comparison_resnet9} shows that, \textit{ArchA}, achieves an inference accuracy of \unit[86.648]{\%}/\unit[65.061]{\%} on the CIFAR10/CIFAR100 dataset, with an accuracy drop of only \unit[0.12]{\%}/\unit[0.4]{\%} compared to an \gls{fp16}-based near-memory logic.

To enhance the consistency of our study, the \gls{nmpu} based on \textit{ArchA} has been integrated into the software stack of the hardware platform proposed by~\cite{Khaddam_2022}, enabling chip-in-the-loop simulation.
The existing \gls{fp16}-based near-memory logic has been disabled from the physical chip. The chip is linked with the software environment where the \gls{nmpu} is implemented through a communication infrastructure that allows them to interact for the computation of subsequent layers.
\cref{fig:accuracy_comparison_resnet9} shows the results for ResNet9 trained on CIFAR10 dataset when using a real hardware platform (marked with a star in the figure). 
As the previous simulations did not account for all chip non-idealities, the real-hardware experiments show an accuracy degradation compared to software simulations.
Nonetheless, the proposed fixed-point \gls{nmpu} implementation achieves competitive performance compared to the \gls{fp16}-based system, consistent with the simulation results.
To fully exploit the benefits of adopting a flexible fixed-point-based architecture, in the next section, we detail its physical implementation and \gls{ppa} assessment.

\section{Synthesis of NMPU in 14nm CMOS technology}
\begin{figure}[t]
\centerline{\includegraphics[width=0.58\columnwidth]{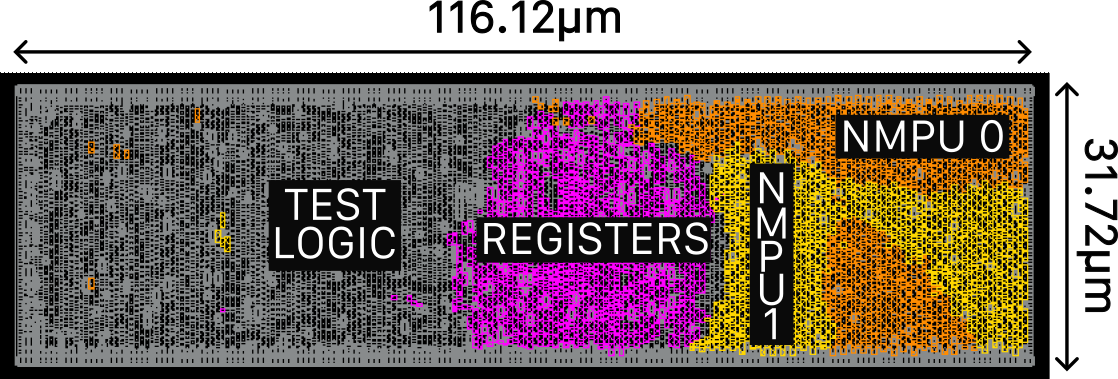}}
\caption{Micrograph of the implemented block in 14nm.}
\label{fig:pnr_results}
\end{figure}

We synthesize \textit{ArchA} using Cadence Genus 20.11 and perform physical implementation using Cadence Innovus 21.13 in a \unit[14]{nm} CMOS technology. 
The resulting \gls{ppa} estimations are shown in \cref{table:pnr_estimation}.
The \gls{nmpu} results in a \unit[505]{$\times$} area reduction compared to \gls{fp16}~\cite{Khaddam_2022}, whose area is estimated to be about \unit[1.67]{MGE} (Gate Equivalent). 
The significant area reduction enables to time-multiplex the \gls{nmpu} across 4 columns of architectures with an \gls{adc} pitch of \unit[4]{$\mu$m}, e.g.~\cite{Khaddam_2022, Gallo_2023}.
Therefore, the lean design allows fitting 64 \glspl{nmpu} operating in parallel in such systems.
The resulting latency and power consumption for post-processing of an \gls{aimc} core~\cite{Khaddam_2022} read \unit[4]{ns} and \unit[24.5]{mW}, respectively.  
This implies that the proposed approach is \unit[139]{$\times$} faster than \gls{fp16}~\cite{Khaddam_2022}, which requires about \unit[558]{ns} to process all the 256 outputs in a serial-pipelined fashion.
The micrograph of the unit, including two fixed-point-based \glspl{nmpu}, the configuration registers for the scale and offset parameters, and the test interface is shown in \cref{fig:pnr_results}.

\begin{table}[t]
\renewcommand{\arraystretch}{1.3}
\caption{Post-layout metrics in 14nm}
\label{table:pnr_estimation}
\centering
\setlength{\tabcolsep}{3.5pt} 
\renewcommand{\arraystretch}{1.3} 
\resizebox{0.9\columnwidth}{!}{
\begin{tabular}{c|ccccc}
\multirow{2}{*}{\makecell{\textbf{System}}} & \multirow{2}{*}{\makecell{\textbf{Area}\\ \textbf{(kGE)}}} & \multirow{2}{*}{\makecell{\textbf{Latency}\\ \textbf{(ns)}}} 
& \multirow{2}{*}{\makecell{\textbf{Tot Latency} \\\textbf{(ns)}}} &\multicolumn{2}{c}{\textbf{Power (mW)}} \\
& & & & \textbf{SS} & \textbf{FF}\\
\hline 
\textbf{Arch}\textit{\textbf{A}} & 3.3 & 1 & 256 & 0.383 & 0.524\\
\makecell{\textbf{Arch}\textit{\textbf{A}}\textbf{$\times$64} } & 211 & 1 & 4 & 24.5 & 33.5\\
\makecell{\textbf{\gls{fp16}} \cite{Khaddam_2022}} & 1666 & 46 & 558 & \multicolumn{2}{c}{27 (post-layout=32)}\\
\end{tabular}
}
\end{table}

\section{Conclusion}
In this paper, we proposed a fixed-point \gls{nmpu} designed, simulated, and physically implemented using \unit[14]{nm} CMOS technology.
Our \gls{nmpu} achieves competitive precision on both small and large networks, i.e., ResNet9 and ResNet32, with accuracy drops of only \unit[0.12]{\%} and \unit[0.4]{\%} compared to existing \gls{aimc} solutions adopting \gls{fp16} precision while minimizing the area overhead by \unit[505]{$\times$}.
Furthermore, due to \gls{nmpu}'s compact form factor, we proved its high-parallelization capability by fitting $\times$64 fixed-point based \glspl{nmpu} in a \unit[7.8]{$\times$} smaller area than the \gls{fp16}-based implementation of~\cite{Khaddam_2022}.
This results in a \unit[139]{$\times$} speed-up for the whole post-processing computation without increasing the power consumption.

\section{Acknowledgments}
We thank M. Ishii and M. Br\"andli for technical support. This work was supported by IBM Research AI Hardware Center, European Union’s Horizon Europe Research and Innovation Program (Grant 101070634), and Swiss State Secretariat for Education, Research and Innovation (SERI) (Grant 23.00205).

\vspace{12pt}

\end{document}

%% file: ieee_cols.tex
\definecolor{ieee-bright-dblue-100}{rgb}{0.0, 0.3828, 0.6055}
\definecolor{ieee-bright-dblue-80}{rgb}{0.0, 0.4883, 0.6797}
\definecolor{ieee-bright-dblue-60}{rgb}{0.3633, 0.6094, 0.7617}
\definecolor{ieee-bright-dblue-40}{rgb}{0.5898, 0.7383, 0.8398}
\definecolor{ieee-bright-dblue-20}{rgb}{0.8906, 0.8984, 0.9219}
\definecolor{ieee-bright-red-100}{rgb}{0.7266, 0.0469, 0.1836}
\definecolor{ieee-bright-red-80}{rgb}{0.832, 0.3164, 0.3281}
\definecolor{ieee-bright-red-60}{rgb}{0.8906, 0.4922, 0.4805}
\definecolor{ieee-bright-red-40}{rgb}{0.9336, 0.6562, 0.6406}
\definecolor{ieee-bright-red-20}{rgb}{0.9688, 0.8203, 0.8125}
\definecolor{ieee-bright-orange-100}{rgb}{0.9961, 0.6367, 0.0}
\definecolor{ieee-bright-orange-80}{rgb}{0.9844, 0.6953, 0.3125}
\definecolor{ieee-bright-orange-60}{rgb}{0.9883, 0.7695, 0.4844}
\definecolor{ieee-bright-orange-40}{rgb}{0.9922, 0.8359, 0.6562}
\definecolor{ieee-bright-orange-20}{rgb}{0.9961, 0.9219, 0.8164}
\definecolor{ieee-bright-yellow-100}{rgb}{0.9961, 0.8164, 0.0}
\definecolor{ieee-bright-yellow-80}{rgb}{0.9961, 0.8477, 0.2148}
\definecolor{ieee-bright-yellow-60}{rgb}{0.9961, 0.875, 0.4492}
\definecolor{ieee-bright-yellow-40}{rgb}{0.9961, 0.9062, 0.6328}
\definecolor{ieee-bright-yellow-20}{rgb}{0.9961, 0.9531, 0.8125}
\definecolor{ieee-bright-lgreen-100}{rgb}{0.4688, 0.7422, 0.125}
\definecolor{ieee-bright-lgreen-80}{rgb}{0.5742, 0.7852, 0.332}
\definecolor{ieee-bright-lgreen-60}{rgb}{0.6875, 0.8398, 0.5039}
\definecolor{ieee-bright-lgreen-40}{rgb}{0.793, 0.8906, 0.6641}
\definecolor{ieee-bright-lgreen-20}{rgb}{0.8945, 0.9414, 0.8281}
\definecolor{ieee-bright-dgreen-100}{rgb}{0.0, 0.5156, 0.2383}
\definecolor{ieee-bright-dgreen-80}{rgb}{0.1641, 0.6055, 0.3867}
\definecolor{ieee-bright-dgreen-60}{rgb}{0.3906, 0.6953, 0.5234}
\definecolor{ieee-bright-dgreen-40}{rgb}{0.6094, 0.8008, 0.6719}
\definecolor{ieee-bright-dgreen-20}{rgb}{0.8047, 0.8945, 0.8359}
\definecolor{ieee-bright-purple-100}{rgb}{0.5938, 0.1133, 0.5898}
\definecolor{ieee-bright-purple-80}{rgb}{0.6992, 0.3281, 0.668}
\definecolor{ieee-bright-purple-60}{rgb}{0.7812, 0.4961, 0.7461}
\definecolor{ieee-bright-purple-40}{rgb}{0.8555, 0.6602, 0.8281}
\definecolor{ieee-bright-purple-20}{rgb}{0.9219, 0.8281, 0.9023}
\definecolor{ieee-bright-lblue-100}{rgb}{0.0, 0.6094, 0.6484}
\definecolor{ieee-bright-lblue-80}{rgb}{0.0, 0.6797, 0.7188}
\definecolor{ieee-bright-lblue-60}{rgb}{0.2109, 0.75, 0.7812}
\definecolor{ieee-bright-lblue-40}{rgb}{0.5469, 0.8242, 0.8438}
\definecolor{ieee-bright-lblue-20}{rgb}{0.7695, 0.918, 0.9219}
\definecolor{ieee-bright-cyan-100}{rgb}{0.0, 0.707, 0.8828}
\definecolor{ieee-bright-cyan-80}{rgb}{0.0, 0.7227, 0.9453}
\definecolor{ieee-bright-cyan-60}{rgb}{0.2656, 0.7812, 0.957}
\definecolor{ieee-bright-cyan-40}{rgb}{0.5547, 0.8438, 0.9688}
\definecolor{ieee-bright-cyan-20}{rgb}{0.7773, 0.9141, 0.9805}
\definecolor{ieee-bright-white-100}{rgb}{0.9961, 0.9961, 0.9961}
\definecolor{ieee-bright-white-80}{rgb}{0.9961, 0.9961, 0.9961}
\definecolor{ieee-bright-white-60}{rgb}{0.9961, 0.9961, 0.9961}
\definecolor{ieee-bright-white-40}{rgb}{0.9961, 0.9961, 0.9961}
\definecolor{ieee-bright-white-20}{rgb}{0.9961, 0.9961, 0.9961}
\definecolor{ieee-dark-red-100}{rgb}{0.5234, 0.1211, 0.2539}
\definecolor{ieee-dark-red-80}{rgb}{0.6445, 0.2812, 0.3828}
\definecolor{ieee-dark-red-60}{rgb}{0.7422, 0.4727, 0.5234}
\definecolor{ieee-dark-red-40}{rgb}{0.832, 0.6445, 0.6758}
\definecolor{ieee-dark-red-20}{rgb}{0.918, 0.8203, 0.832}
\definecolor{ieee-dark-orange-100}{rgb}{0.9062, 0.4648, 0.1328}
\definecolor{ieee-dark-orange-80}{rgb}{0.9648, 0.5664, 0.3164}
\definecolor{ieee-dark-orange-60}{rgb}{0.9766, 0.6758, 0.4805}
\definecolor{ieee-dark-orange-40}{rgb}{0.9844, 0.7773, 0.6523}
\definecolor{ieee-dark-orange-20}{rgb}{0.9922, 0.8789, 0.8125}
\definecolor{ieee-dark-yellow-100}{rgb}{0.9961, 0.7773, 0.1719}
\definecolor{ieee-dark-yellow-80}{rgb}{0.9961, 0.8086, 0.375}
\definecolor{ieee-dark-yellow-60}{rgb}{0.9961, 0.875, 0.4492}
\definecolor{ieee-dark-yellow-40}{rgb}{0.9961, 0.8984, 0.6875}
\definecolor{ieee-dark-yellow-20}{rgb}{0.9961, 0.9453, 0.8438}
\definecolor{ieee-dark-lgreen-100}{rgb}{0.3945, 0.5508, 0.0938}
\definecolor{ieee-dark-lgreen-80}{rgb}{0.5078, 0.6289, 0.293}
\definecolor{ieee-dark-lgreen-60}{rgb}{0.6367, 0.7188, 0.4688}
\definecolor{ieee-dark-lgreen-40}{rgb}{0.7539, 0.8047, 0.6367}
\definecolor{ieee-dark-lgreen-20}{rgb}{0.875, 0.9023, 0.8125}
\definecolor{ieee-dark-dgreen-100}{rgb}{0.0, 0.3867, 0.2539}
\definecolor{ieee-dark-dgreen-80}{rgb}{0.1836, 0.5, 0.3906}
\definecolor{ieee-dark-dgreen-60}{rgb}{0.3984, 0.6172, 0.5273}
\definecolor{ieee-dark-dgreen-40}{rgb}{0.5938, 0.7422, 0.6758}
\definecolor{ieee-dark-dgreen-20}{rgb}{0.793, 0.8711, 0.8359}
\definecolor{ieee-dark-purple-100}{rgb}{0.4648, 0.1445, 0.5117}
\definecolor{ieee-dark-purple-80}{rgb}{0.5898, 0.3242, 0.6016}
\definecolor{ieee-dark-purple-60}{rgb}{0.6914, 0.4883, 0.6953}
\definecolor{ieee-dark-purple-40}{rgb}{0.7969, 0.6523, 0.793}
\definecolor{ieee-dark-purple-20}{rgb}{0.8945, 0.8203, 0.8945}
\definecolor{ieee-dark-cyan-100}{rgb}{0.0, 0.4492, 0.4648}
\definecolor{ieee-dark-cyan-80}{rgb}{0.0, 0.5469, 0.5664}
\definecolor{ieee-dark-cyan-60}{rgb}{0.3047, 0.6602, 0.668}
\definecolor{ieee-dark-cyan-40}{rgb}{0.5586, 0.7695, 0.7734}
\definecolor{ieee-dark-cyan-20}{rgb}{0.7734, 0.8789, 0.8789}
\definecolor{ieee-dark-dblue-100}{rgb}{0.0, 0.1562, 0.332}
\definecolor{ieee-dark-dblue-80}{rgb}{0.1797, 0.3008, 0.4609}
\definecolor{ieee-dark-dblue-60}{rgb}{0.3828, 0.4609, 0.5859}
\definecolor{ieee-dark-dblue-40}{rgb}{0.5781, 0.6289, 0.7188}
\definecolor{ieee-dark-dblue-20}{rgb}{0.7852, 0.8047, 0.8555}
\definecolor{ieee-dark-grey-100}{rgb}{0.457, 0.4688, 0.4805}
\definecolor{ieee-dark-grey-80}{rgb}{0.5625, 0.5625, 0.5742}
\definecolor{ieee-dark-grey-60}{rgb}{0.6641, 0.6641, 0.6758}
\definecolor{ieee-dark-grey-40}{rgb}{0.7734, 0.7695, 0.7773}
\definecolor{ieee-dark-grey-20}{rgb}{0.8789, 0.8828, 0.8828}
\definecolor{ieee-dark-black-100}{rgb}{0.0, 0.0, 0.0}
\definecolor{ieee-dark-black-80}{rgb}{0.3438, 0.3477, 0.3555}
\definecolor{ieee-dark-black-60}{rgb}{0.5, 0.5078, 0.5195}
\definecolor{ieee-dark-black-40}{rgb}{0.6523, 0.6602, 0.6719}
\definecolor{ieee-dark-black-20}{rgb}{0.8164, 0.8242, 0.8281}
\definecolor{ieee-bright-emerald-100}{rgb}{0.459, 0.925, 0.624}